# The development of an electrochemical sensor for antibiotics in milk based on machine learning algorithms

Timur A. Aliev[†], Vadim E. Belyaev[†], Anastasiya V. Pomytkina, Pavel V. Nesterov, Sergei V. Shityakov, Roman V. Sadovnychiy, Alexander S. Novikov, Olga Yu. Orlova, Maria S. Masalovich, Ekaterina V. Skorb*

*ITMO University, Lomonosova str. 9, 191002 Saint-Petersburg, Russia*

*Correspondence to: skorb@itmo.ru

**Present study is dedicated to the problem of electrochemical analysis of multicomponent mixtures such as milk. A combination of cyclic voltammetry facilities and machine learning technique made it possible to create a pattern recognition system for antibiotic residues in skimmed milk. A multielectrode sensor including copper, nickel and carbon fiber was fabricated for the collection of electrochemical data. Chemical aspects of processes occurring at the electrode surface were discussed and simulated with the help of molecular docking and density functional theory modelling. It was assumed that the antibiotic fingerprint reveals as potential drift of electrodes owing to redox degradation of antibiotic molecules followed by *pH* change or complexation with ions present in milk. Gradient boosting algorithm showed the best efficiency towards training the machine learning model. High accuracy was achieved for recognition of antibiotics in milk. The elaborated method may be incorporated into existing milking systems at dairy farms for monitoring the residue concentrations of antibiotics.**

In accord with arising problems of the Earth population growth and globalization of the society, it is obligatory to maintain standards of living and food quality. On the other hand, keeping up with time is getting impossible without robotization, automatization and artificial intelligence[1-5]. At present, agriculture and food industry are focused on implementing electronic sensing devices such as e-tongues and e-noses providing timely analysis of food and decreasing risks of its deterioration, contamination, and adulteration[6-16]. Electronic devices employing electrochemical sensors attract frontier research in biology, electrochemistry, and electrochemical engineering. Novel developments make it possible to detect a wide spectrum of organic and inorganic molecules as well as bacteria and viruses[17-19]. Meanwhile, modern research in artificial sensing needs to cover a larger scope of objects. Despite their primary role of pharmaceuticals, antibiotics are subtle components of food and milk which causes a serious threat for the human health and environment.

Antibiotics, as the only substantially effective drugs against bacterial infections, are indispensable for the treatment of human and animal diseases. On the other hand, their uncontrollable usage provokes a serious contamination of the environment and increases the resistance of pathogenic bacteria[20,21]. Antibiotics are commonly added to feedstuff for domestic animals. The molecules of most antibiotics are difficult to completely decompose in the body, and some quantities of the toxic substances are eventually spread unaltered to the human food production chain (milk and meat). The danger of antibiotics application is commonly associated with such side-effects as allergic reaction, dysbacteriosis and even nervous system damage[22]. Due to the necessity to minimize risks of food contamination, authorities worldwide enact maximum residue limits (MRLs) for veterinary antibiotics. Thus, it is of utmost importance to pave the way towards break-through technologies facilitating precise, prompt and point-of-use control of the level of the residues.

Nowadays, there are a few conventional precise analytical methods for determination of antibiotics such as mass-spectrometry and high-performance liquid chromatography with different modes of detection. However, they are long lasting, expensive, and often too laborious, when some procedures such as derivatization, extraction and purification are included[23-25].

Bioanalytical assays including lateral flow immunochromatography (ELISA, enzyme-linked immunosorbent assay) in the form of test strips have already gained commercial success owing to availability, portability, simplicity, and low cost. But this methodology is limited to only qualitative analysis[24,26].

In recent decades, electrochemical sensors based on voltammetric techniques, chronoamperometry and electrochemical impedance spectroscopy have shown enormous potential towards the qualitative and quantitative detection of antibiotic residues[22,24,25,27].

Yet, on analyzing multicomponent biological liquids like milk electrochemically, it may be difficult to unambiguously categorize multiple signals. This obstacle can be overcome by collecting and afterward statistic processing big data blocks containing such a response via an algorithm of machine learning.

In this study we combine a classical electrochemical method with opportunities of machine learning (ML) technique which has already proved to be effective and promising[23,28-36]. Machine learning is a technology that allows a computer to independently acquire information from data. It enables a computer system to respond to an input based on the optimization (through a large amount of data and computing power) of a statistical model and make predictions with reasonable accuracy[37,38]. Here we introduce a novel electrochemical sensor for both qualitative and quantitative analysis of an antibiotic series in milk where sensing is based on cyclic voltammetry response referring to a statistical model obtained as a result of machine learning of a big data set[39].

**Results**

The elaborated procedure of antibiotic recognition in milk is depicted in Figure 1. At first step the electrochemical sensor representing a multielectrode system is introduced to a solution of milk (Figure 1 a-b.) Three subsequent cyclic voltammograms (CVs) are registered for a sample of milk (Figure 1 c). The obtained set of discrete values ($dI/dE$) is processed via referring to the statistical model based on the previously collected database, and the prediction is displayed on the interface of a computer (Figure 1 d-f). For collecting the data base of 5 antibiotics and blank milk we registered a total number of 1377 CVs which were then processed using a gradient boosting algorithm of machine learning (Figure 1 e).

The approach towards multielectrode sensors has been described elsewhere before [7,10,40,41]. It has been shown that multisensors are promising for acquiring complex analytical signals for advanced data processing with chemometric techniques. The multisensors are commonly arrays of relatively non-specific electrodes that respond to analytes. The sensing signal generated from each sensor is complementary, and the combination of them generates a unique "fingerprint". Utilizing this concept, we employed three conductive materials: copper, nickel, and carbon fiber, which were duplicated to obtain an integrated working electrode. The final equivalent circuit of a two-electrode electrochemical cell is shown in Figure 1 b.

**Electrochemical studies of individual electrodes in blank milk and antibiotic solutions**

To elucidate processes occurring at the multielectrode system in presence or absence of an antibiotic, each working electrode was investigated separately in pair with a copper counter electrode. Figure 2 a-c features responses from three different two-electrode electrochemical cells: Cu-Cu, Ni-Cu and C-Cu, respectively. In general, the shape of CVs within each figure is highly reproducible, independently on the presence of antibiotics, and is presumably a characteristic of electrode materials.

The observed plateau at $+0.3 \div +1.2$ V as well as peak in the range $0 \div -0.5$ V for the Cu-Cu cell (Figure 2a) can be attributed to copper-working and copper-counter electrode oxidation from metallic Cu or other species in accordance with the equations of half-reactions[42]:

$$Cu_2O\ (s) + 2e^- + 2H^+ = 2Cu^0\ (s) + H_2O \tag{1}$$

$$2Cu^{2+} + 2e^- + H_2O = Cu_2O\ (s) + 2H^+ \tag{2}$$

$$Cu^{2+} + 2e^- = Cu^0\ (s) \tag{3}$$

$$2Cu(OH)_2\ (s) + 2H^+ + e^- = Cu_2O\ (s) + 3H_2O \tag{4}$$

The reaction (equation 3) is less favorable as it requires a very strong acidic medium with a *pH* of lower than 3.0.

Despite the probability of the reaction (4) at milk *pH* of about 6.7÷6.9[43], XRD analysis does not prove the formation of copper (II) hydroxide (see XRD spectra, Figure 2e).

The oxidation peaks for Ni-Cu system (Figure 2b) may be associated with the formation of $Ni^{2+}$ (equation 5), nickel oxyhydroxide (equation 6) and $NiO_2$ (equation 7, 8) species[44]:

$$Ni^{2+} + 2e^- = Ni^0 \text{ (s)} \tag{5}$$

$$NiOOH \text{ (s)} + 3H^+ + e^- = Ni^{2+} + 2H_2O \tag{6}$$

$$NiO_2 \text{ (s)} + 4H^+ + 2e^- = Ni^{2+} + 2H_2O \tag{7}$$

$$NiO_2 \text{ (s)} + 2H_2O + 2e^- = Ni(OH)_2 \text{ (s)} + 2OH^- \tag{8}$$

Carbon cloth is an inert material but rather absorptive. Thus, the anodic and cathodic curves may be attributed to absorbance of milk species or antibiotic molecules at the working electrode as well as oxidation processes at the copper-counter electrode (Figure 2c). Despite the fact, that bubbles of gases were not observed, the start of water decomposition cannot be completely excluded[45,46]:

$$O_2 + 4H^+ + 4e^- = 2H_2O \tag{9}$$

$$2H_2O + 2e^- = H_2 + 2OH^- \tag{10}$$

The peculiar feature of all CVs registered for milk solutions containing antibiotics is the shift of potentials to the greater values and increase in anodic and cathodic currents. Considering these reactions *pH*-dependent, we may presume that the shift is caused by the increase of acidity in milk due to antibiotic destruction.

Tetracyclines are known to undergo electrolysis in the positive range of potentials producing protons[47]. Streptomycin is also shown to oxidize irreversibly with the formation of streptomycinic acid which acidic nature also contributes to the overall acidity[48]. The increase in proton concentration shifts the equilibrium of *pH*-dependent reactions towards reduced forms (equations 1,4,6-10) which in turn results in the rise of redox potentials $E$ according to a Nernst equation brought for room temperature (T=298 K) (equation 11)[49-52]:

$$E = E^o - 0.059v/n(pH) \tag{11}$$

where $E^o$ – a standard redox potential for the given half-reaction, $v$ – a stoichiometric coefficient for hydrogen protons in the equation of a half-reaction, $n$ – a number of electrons participating in the reaction, and $pH = -\lg[\alpha_{H+}]$, where $[\alpha_{H+}]$ – equilibrium concentration of active hydrogen species.

Considering the whole scope of physical and chemical pathways for reactions at the electrode-electrolyte interface, it should be mentioned that antibiotic molecules such as tetracyclines, penicillins, cephalosporins and aminoglycosides represent ligands with moieties which may effectively coordinate to copper (2+) and nickel (2+) ions forming chelate compounds[48,53-55].

Tetracycline molecules exhibit several potential metal-binding sites: oxygens at the $C_{10}$-$C_{12}$ phenolic β-diketone system, enolic oxygens, and the nitrogens at $C_4$ and at the carboxamide group in ring A[53].

For Cu (II)-TC the stoichiometry of the complexes (metal ion/ligand) were found to be 1:1[54]. Relying on the results of XRD-analysis (see XRD and EDX-analysis), we can suggest copper (I) chloride undergoing a reaction with a tetracycline molecule via disproportionation with the formation of a charged six-coordinated copper (II) complex (Figure 2e) (equation 12):

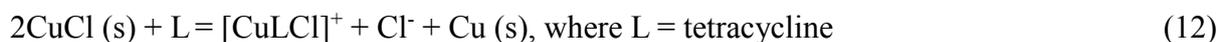

$$2CuCl\ (s) + L = [CuLCl]^+ + Cl^- + Cu\ (s),\ where\ L = tetracycline \tag{12}$$

Making a presumption that reaction (12) is taking place at copper surface, the shift of the electrode potential will depend on chloride ions equilibrium concentration, solubility product constant of copper (I) chloride, and stability constant of $[CuLCl]^+$- complex.

For penicillins, the coordination in Ni (II) complex involves the O atoms of the beta-lactam group (Figure 2f)[55].

Studies of solid complexes of cefazoline with Ni(II) and Cu(II) led to the conclusion that Czl⁻ is coordinated via the carboxylate and amide group O atom and the N atom of the heterocyclic side chain (Figure 2g).

Molecules of tetracyclines, penicillins, cephalosporins and some aminoglycosides form thermodynamically effective and stable complexes[48,53-55]. Streptomycin, on the contrary, reveals a tendency to hydrolyze in the presence of metal ions[48]. This phenomenon supports the idea that copper and nickel ions as well as other species of copper such as $Cu^0$ and copper oxides can catalyze degradation of streptomycin molecules via formation of intermediate complexes followed by cleavage of covalent bonds (Figure 2h)[56,57].

**XRD and EDX-analysis**

The XRD spectra of copper plates exposed to electrochemical cycling in a solution of blank milk unambiguously show the dominant presence of metallic copper (Figure 2d). Peaks of significantly lower intensity in comparison with those for $Cu^0$ are associated with copper (I) chloride and copper (I) oxide.

The EDX-analysis of copper plates reveals a series of elements adsorbed on the surface of the electrode such as carbon, potassium, calcium, oxygen, phosphorous and chlorine. In fact, in milk, all these macro-elements are distributed differently into diffusible and non-diffusible fractions[43]. Potassium and chloride ions are essentially diffusible although calcium and inorganic phosphate are partly bound to the casein micelles. Analyzing the EDX-spectrum of copper electrodes we obviously deal with aqueous phase because small ions are much more mobile under the influence of external electric field than heavy globules of proteins forming colloidal particles. The significant percentage of carbon (75.32 at. %) and oxygen (8.70 at. %) may be attributed to organic soluble components - lactose, albumin, and globulin. The EDX-spectra do not contradict with the presence of insoluble copper (I) oxide, which is the product of redox processes (equations 1,2). Copper (I) chloride is possibly generated via the reduction of $Cu^{2+}$ (equation 13) or disproportionation reaction (equation 14) in the presence of chlorine ions.

$$Cu^{2+} + Cl^- + e^- = CuCl\ (s) \tag{13}$$

$$Cu_2O\ (s) + 2Cl^- + 2H^+ = 2CuCl\ (s) + H_2O \tag{14}$$

**Molecular docking modelling**

Considering the multicomponent system complicated, we presumed that electrode processes are not strictly diffusive and can be accompanied by adsorption phenomenon. To study a possible pathway towards formation of intermediate complexes of antibiotic molecules with substrate materials, we designed their models predicting 3d structures (Figure 2a-c) and calculated free Gibbs energies with the help of AutoDock Vina. Calculations showed that all the structures with antibiotics are characterized with small but negative values of binding energies which indicates moderate affinities and thermodynamic efficiency. For the tetracycline molecule the greatest affinity was found towards nickel (-7.5 kcal/mol), and the least towards carbon (-6.8 kcal/mol).

**Density functional theory (DFT) modelling**

We also carried out quantum chemical calculations (see Table S1–S2) in order to compare thermodynamic favorability for the supramolecular association of penicillin V and streptomycin molecules with Cu and $Cu_2O$ model species which were revealed in course of XRD-analysis (Figure 2d). The results of DFT-modelling are shown in Figure 2i. The supramolecular association of penicillin V with $Cu_2O$ was found more thermodynamically favorable compared with appropriate concurrent self-assembly process with Cu (by 40.3 kcal/mol in terms of Gibbs free energies), whereas, in case of streptomycin molecule, situation is opposite: the supramolecular association of streptomycin molecule with Cu is 7.0 kcal/mol more thermodynamically beneficial in terms of Gibbs free energies. Thus, results of our quantum chemical calculations reveal that penicillin V and streptomycin molecules should selectively adsorb on $Cu_2O$ and Cu surfaces, respectively.

**Machine learning**

The binary classification model shows good accuracy of 0.99 for the antibiotic class and 0.74 for the milk class. This implies that antibiotics are recognized more effectively while milk is confused with antibiotics in 26 cases of 100 (Figure 3a). Such a result is obviously explained by the nonuniformity of the database as the antibiotic set is formed of 5 times bigger data array than the milk set. The receiver operating characteristic curve (ROC) is above the chance line and quickly transition from the origin towards (0.1), thus being characterized by a high true positive rate (TPR) and a small false positive rate (FPR), which indicates a good prediction strategy. Precision-recall curve is another qualitative tool to assess the performance of the trained model. It demonstrates a trade-off between accuracy (recall) and false positive rate (precision) for milk and antibiotic classes showing superiority of the last one.

The same data set used for the binary classification was applied in multiple classification where each antibiotic was represented as a single class (Figure 3b). In general, high values of the diagonal matrix indicate the correct operation of the trained model. The performance of milk class is better in comparison with the binary classification model (0.8 versus 0.74) owing to the increased uniformity of the data set. It can be seen from the confusion matrix as well as ROC and precision-recall curves that the tetracycline class reveals the best characteristics while the ceftiofur and milk ones the worst.

A regression model is another supervised learning algorithm utilized for prediction of the labelled datasets. The main characteristic of the linear regression model is the coefficient of determination ($R^2$) which can be interpreted as the difference between the samples in the dataset and the predictions made by the model. However, quite a low value of 0.9394 lets us assert the model to be less effective than the classification one (Figure 3c).

Figure 3d demonstrates the application of multiclassification model in an altered dataset containing more classes but less data for each class. It can be seen from the confusion matrix that the accuracy of prediction has significantly fallen, obviously due to the lack of features (voltammograms).

The Fedot framework [https://github.com/nccr-itmo/FEDOT][58,59] was utilized as an alternative to CatBoost algorithms showing the same tendencies towards classification and regression but revealing a lower performance (Figure S2-S9).

All data used to train the models are available on the QR code (Figure 3e).

**Discussion**

A complex interdisciplinary study has been accomplished and the novel approach towards detection of a series of antibiotics in skimmed powdered milk has been successfully applied. The approach is founded on a classical electrochemical method (cyclic voltammetry) in combination with an artificial intelligence tool (machine learning) operating with big data. On studying the chemical aspects of interfacial processes at electrodes it was assumed that the antibiotic fingerprint reveals as potential drift of electrodes owing to redox degradation of antibiotic molecules followed by *pH* change or complexation with ions present in milk.

The elaborated procedure may be extended to any other series of antibiotics and become a steppingstone to analytical recognition of a wide spectrum of molecules in complex solutions, dispersions, and mixtures via collecting big data bases.

The proposed assay contributes to the development of precise, prompt, easy-to-use and low-cost methods and is intended to provide a timely analysis of raw milk in field conditions.

On creating digital fingerprint for antibiotics in raw milk varying in fat, the developed analytical set including a sensor, potentiostat, database and software is prospective for integration into common automized somatic health analyzers at dairy farms (Figure 4). In fact, advanced milking systems may include conductivity sensors facilitating detection of the inflammation of mammary glands [https://www.afimilk.com/parlor-automation/, https://www.delaval.com/en-au/discover-our-farm-solutions/milking/delaval-vms-series/, https://www.lely.com/solutions/housing-and-caring/treatmentbox/]. The diagnosis is normally followed by mastitis case routine with antibiotics during which the milk is disposed of as waste. Hence, it is of importance to reveal the period needed for antibiotic excretion and predict the first opportunity of commercial utilization of milk as soon as possible via control of antibiotic concentration. Incorporating the antibiotic sensor in the milking system together with other in-line somatic health analyzers, milk farmers will be able to effectively monitor the level of an antibiotic residue and, thus, diminish the monetary losses.

**Methods**

**Electrochemical methods**

Cyclic voltammetric (CV) measurements were performed in a full electrochemical cell under ambient conditions. To prepare a solution of blank milk, skimmed milk powder was diluted in distilled deionized water to attain a concentration of 100 g/l. Antibiotic powders were then dissolved in the blank milk to obtain five concentrations including one of MRL. The potentiostat (CompactStat.h, Ivium Technologies) was used as a potential setting device. CV characteristics were obtained in a potential range from -0.8 to 1.8 V at a scan rate of 50 mV/s. After every three cycles the sensor was withdrawn from the solution and the surface of electrodes cleaned with abrasive paper and rinsed in distilled water. Each concentration of an antibiotic was examined from 3 to 15 times using an independent sample of milk in accordance with this procedure.

For electrochemical studies of individual electrodes copper, nickel wires and a few strands of carbon cloth were employed as working electrodes separately and a copper plate as a counter electrode. The procedure of electrochemical measurements was as described above.

Concentrations of antibiotics were chosen 10 times higher than of MRL, i. e. $2 \cdot 10^{-6}$ g/ml for streptomycin and $1 \cdot 10^{-7}$ g/ml for tetracycline, respectively.

**Multisensor assembly**

Multisensor (multielectrode sensing pen) consists of copper, nickel wires and carbon fibers sealed in epoxy resin so that only a cross-section is exposed to milk forming an interface (electrode-electrolyte). A copper wire with larger diameter acts as a counter electrode (Figures S10-S12).

**Machine learning**

The arrays of «current per voltage» values of two databases were used to create machine learning models. The first consisted of 6 classes (blank milk, streptomycin, tetracycline, penicillin V, cefazolin and ceftiofur), the second one − of 16 classes (blank milk and 15 antibiotics). An array, received from a single voltammogram, consisted of 1040 values which were decided to take as features. The first database contained 1,432,080 features out of 1377 CVs. The second database contained 2,206,880 features out of 2122 CVs. Also, concentration values and class names were added to arrange all CVs in the databases.

The Python Pandas and Numpy programming language modules were used for data preprocessing. During pre-processing, all CVs were combined into one table and saved in csv format. The resulting database table consisted of an array of CV «currents per voltage», class name, and antibiotic concentration.

The Python scikit-learn programming language modules, CatBoost, matplotlib, and the Fedot framework were used to assess the machine learning models. The ratios of training to testing data: 50 to 50, 80 to 20, and 90 to 10 were selected, after which the best model was selected. The models were estimated on separate test data which did not participate in the training process, respectively. To create a model and optimize its hyperparameters, the CatBoost gradient boosting framework and the built-in method of the grid search model class were used.

Binary classification models (blank milk or antibiotic), multiclassification models (blank milk and a specific antibiotic), as well as regression models were trained. Machine learning pipelines were also created using the Fedot framework.

The Fedot framework was used as a baseline, owing to its ability to create pipelines automatically (automl) and its API for creating pipelines. Pipelines based on automatic machine learning and using the Fedot API for 5 databases with 5 antibiotics, as well as for a database with 16 antibiotics were created (Figure S2-S9).

**XRD analysis**

The samples of copper plates corroded in due course of electrochemical oxidation while registering CVs in milk were investigated with the powder X-ray diffraction (XRD) method (Bruker D2 PHASER, Germany). The experimental conditions were as follows: CuKα irradiation, 2θ range of 10–100°.

For XRD analysis two copper plates were immersed in a solution of blank milk and exposed to linear potential change in the range from -0.8 to 1.8 V for a hundred cycles to collect as much product as possible. Then the electrodes were rinsed in distilled water to wash out all soluble components.

**EDX analysis**

Measurements were conducted with a scanning electron microscope (SEM) (Quanta Inspect instrument, FEI, USA) in vacuum at pressure of about $10^{-3}$ - $10^{-4}$ Pa and an increasing voltage of 30 kV in the secondary electron mode. Qualitative and quantitative elemental analysis was carried out with the use of energy dispersive X-ray spectroscopy (EDX) under nitrogen cooling. The procedure of sample preparation was similar to that of XRD though the samples were not rinsed to show absorbance of elements containing in milk during cycling.

**Molecular docking method**

AutoDock Vina v.1.2.0 was used for all dockings in this study[60]. In general, the docking parameters were kept to their default values. The size of the docking grid was 18 Å × 40 Å × 40 Å for copper, 22 Å × 44 Å × 42 Å for nickel, and 44 Å × 40 Å × 28 Å for carbon. The results of molecular docking were visualized utilizing PyMol 2.5 software.

3D-structures of ligands were downloaded from PubChem database https://pubchem.ncbi.nlm.nih.gov/ and used as received.

**Density functional theory computational details**

The full geometry optimization of all model structures was carried out at the ωB97XD/CEP-121G level of theory with the help of the Gaussian-09 [M. J. Frisch, G. W. Trucks, H. B. Schlegel et al. in Gaussian 09, Revision C.01, Gaussian, Inc., Wallingford, CT, 2010] program package. No symmetry restrictions were applied during the geometry optimization procedure. The Hessian matrices were calculated for all optimized model structures to prove the location of correct minima on the potential energy surface (no imaginary frequencies were found in all cases). The thermodynamic parameters were calculated at 298.15 K and 1.00 atm (Tables S1 and S2). The Cartesian atomic coordinates for all optimized equilibrium model structures are presented in Supplementary Information as xyz-files.

**Supplementary Information**

More results and detailed methods are available in supplementary materials.

**Acknowledgements**

This work was supported by RSF grant № 21-13-00403. We acknowledge Blue Sky Research program. We also highly appreciate contributions of ITMO students, who did collection of dataset, especially, Elena A. Scherbakova, Roman V. Sergienko, Ilya K. Eliseev, Alexey A. Kulchitskiy, Alexey A. Ryazanov and Shirin M. Ramazanova, and other students of Academia Talant and Sirius.Leto Program for dataset collection.

**Figure captions**

**Figure 1**. Schematic route of the antibiotic assay: **a**, voltammetric stand; **b**, a photo of the multielectrode (bottom view) and its composition; scheme depicting equivalent circuit of the two-electrode electrochemical cell; *Z w.e.* and *Z c.e.* are impedances of the interface between the working electrode and solution, and the counter electrode and solution, respectively; *Rs* – electrolyte (solution) resistance; **c**, a typical set of three CVAs; **d**, data processing; **e**, a «gradient boosting» diagram; **f**, a «print screen» with a predicted response.

**Figure 2**. Investigation of processes at the electrode – electrolyte interface: **a-c**, CVs for Cu-Cu, Ni-Cu and C-Cu electrode systems, respectively (second cycles); the insets - docking images and affinity energies between a tetracycline molecule and corresponding conductive supports; **d**, X-ray diffraction and energy-dispersive X-ray spectra for a copper electrode after electrochemical cycling; **e-g**, proposed formulae for Cu (II) and Ni (II) complexes of tetracycline, penicillin and cefazoline; **h**, a proposed scheme for hydrolysis of streptomycin catalyzed with copper (I) oxide; **i**, DFT-calculated free Gibbs energies for optimized equilibrium model structures: I – penicillin-Cu, II – streptomycin-$Cu_2O$, III – streptomycin-Cu; IV – penicillin-$Cu_2O$.

**Figure 3**. CatBoost models validation results: **a**, confusion matrix, ROC curves and precision-recall curves for binary classification; **b**, confusion matrix, ROC curves and precision-recall curves for 5 antibiotics classification; **c**, regression validation of test data; **d**, confusion matrix for 16 antibiotics dataset validation; **e**, QR-code on google drive with all datasets and models.

**Figure 4**. Proposed integrated analyzing system: insets **a**, automized somatic health control, **b**, automized antibiotic residue control.

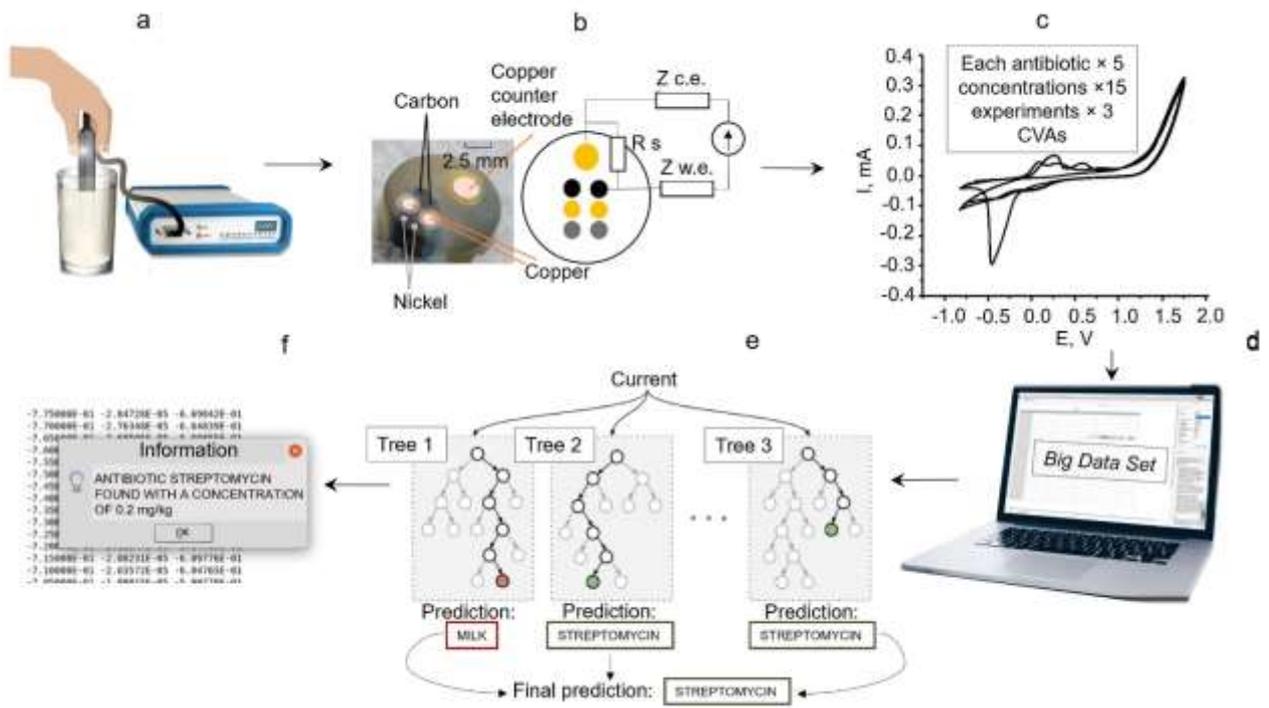

Figure 1.

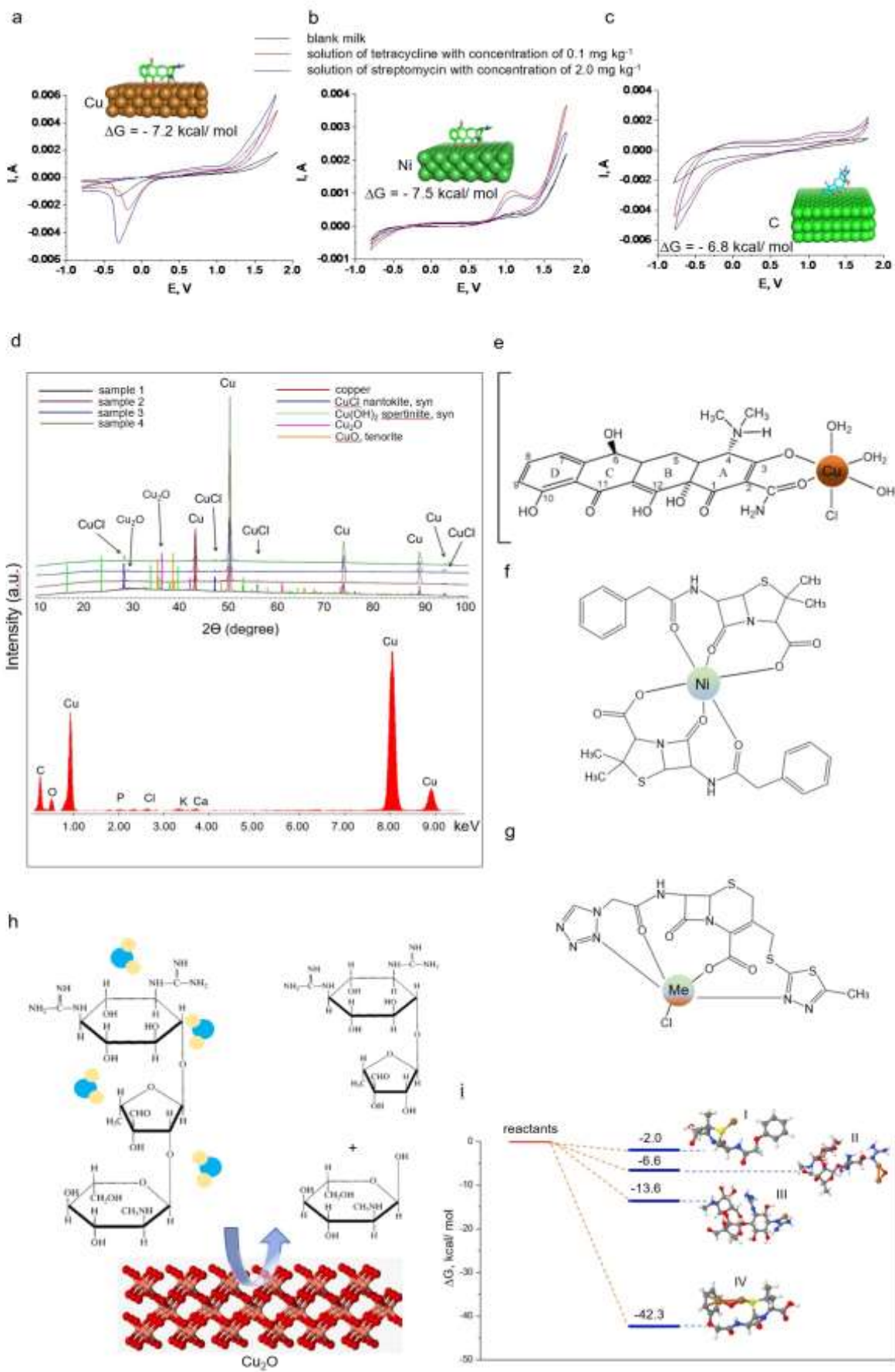

Figure 2.

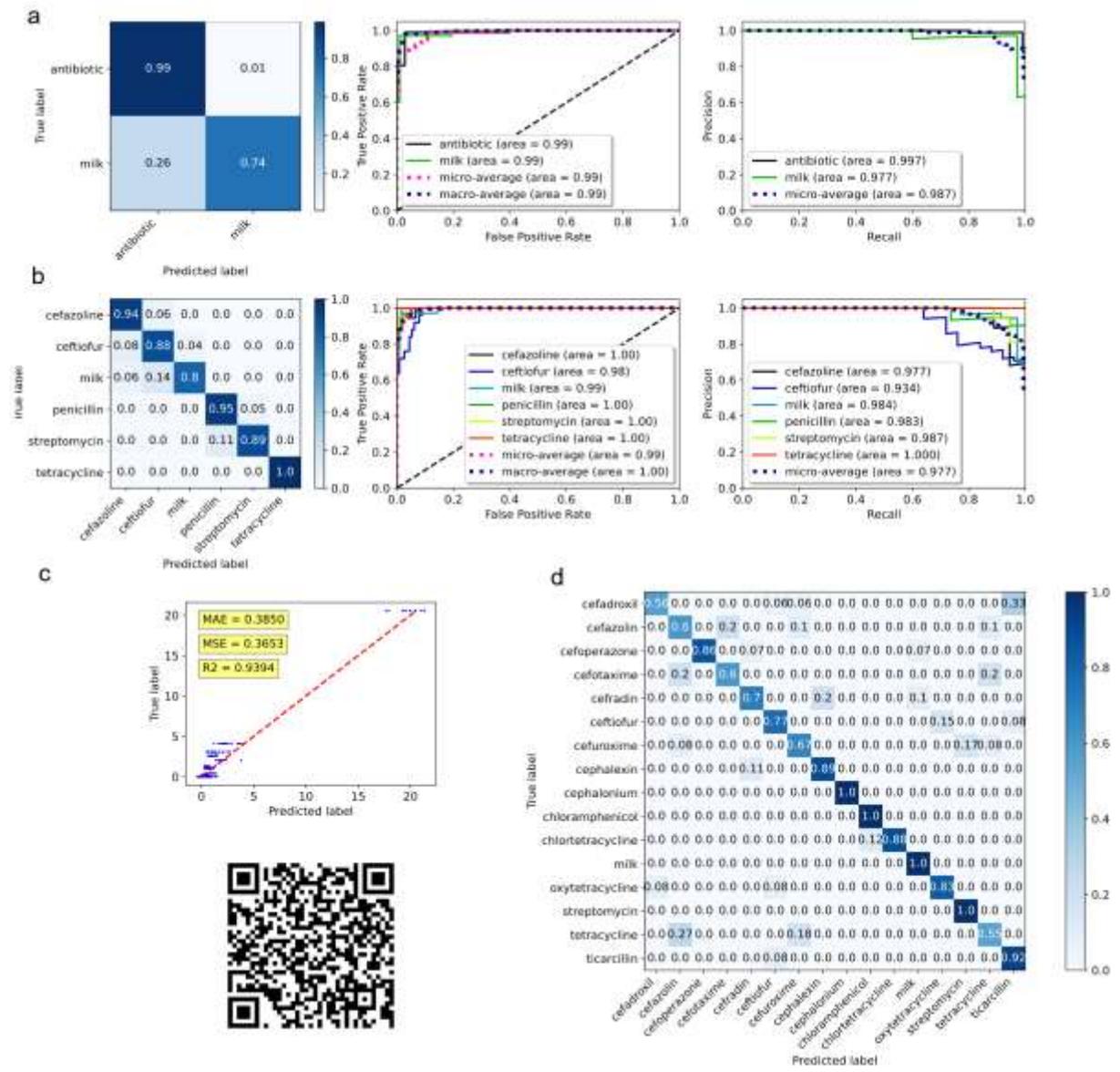

Figure 3.

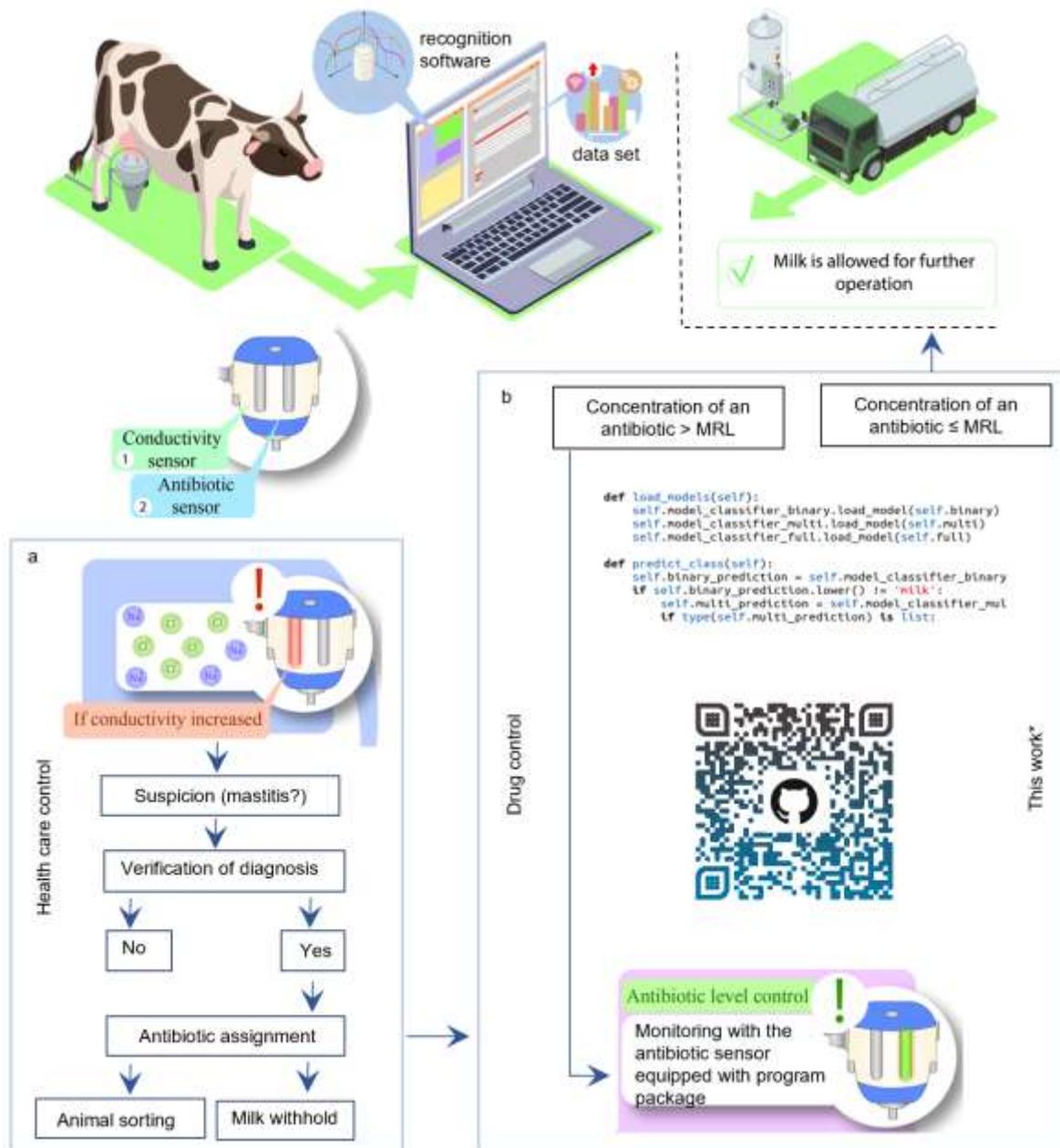

Figure 4.

# Supplementary Information

**The development of an electrochemical sensor for antibiotics in milk based on machine learning algorithms**

Timur A. Aliev[†], Vadim E. Belyaev[†], Anastasiya V. Pomytkina, Pavel V. Nesterov, Sergei V. Shityakov, Roman V. Sadovnychiy, Alexander S. Novikov, Olga Yu. Orlova, Maria S. Masalovich, Ekaterina V. Skorb*

*ITMO University, Lomonosova str. 9, 191002 Saint-Petersburg, Russia*

*Correspondence to: skorb@itmo.ru

**Table of contents:**

## Results

**Electrochemical studies of individual electrodes in blank milk and antibiotic solutions**

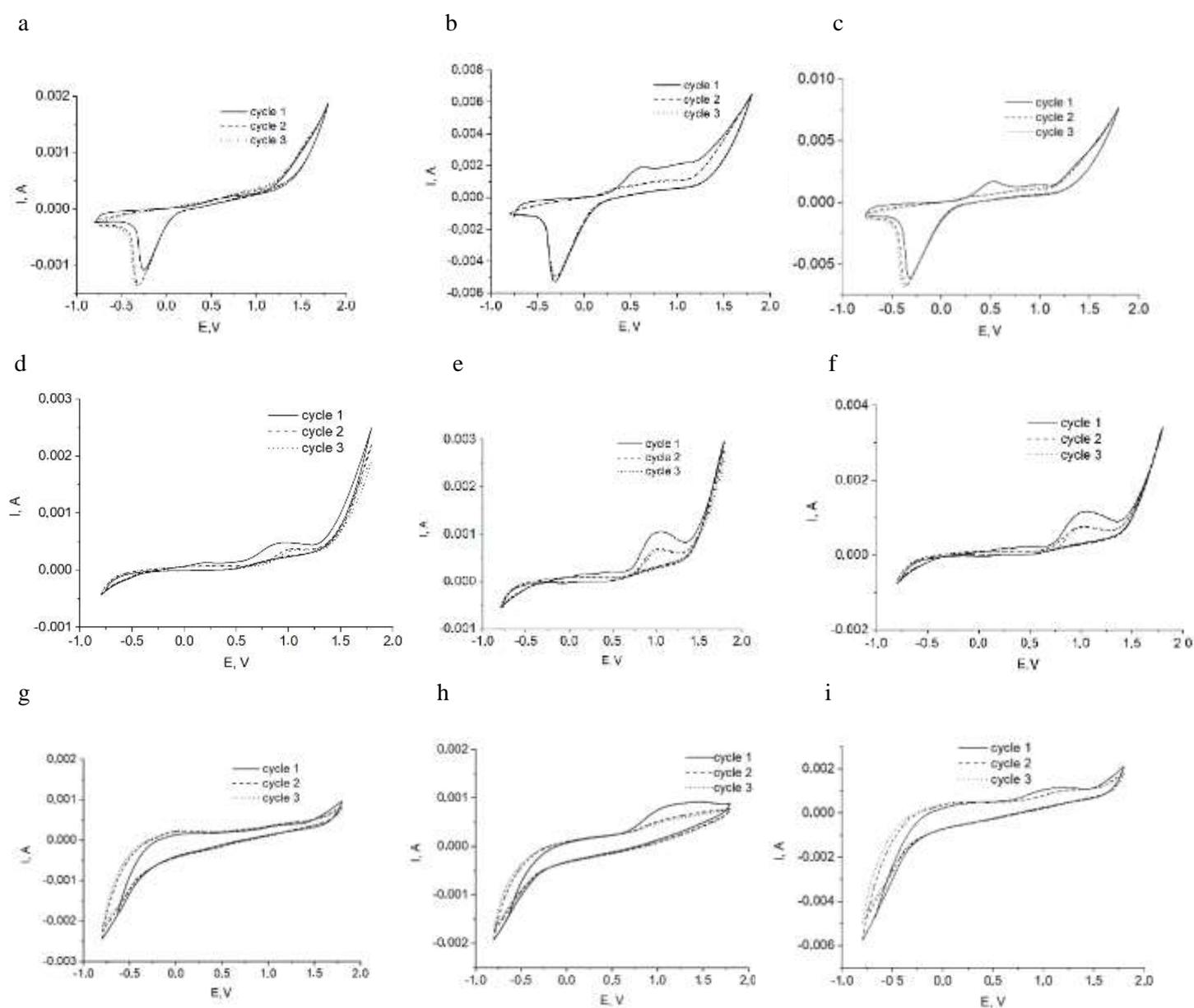

**Figure S1.** Cyclic voltammograms for two-electrode electrochemical cells of Cu-Cu (a, b, c), Ni-Cu (d, e, f) and C-Cu (g, h, i) in blank milk (a, d, g), streptomycin solution in milk (b, e, h) and tetracycline solution in milk (c, f, i).

**Density functional theory (DFT) modelling**

**Table S1**. Calculated enthalpies, entropies, and Gibbs free energies (in Hartree) for optimized equilibrium model structures (H, S, and G, respectively).

| Model structure | H | G | S |
|---|---|---|---|
| Cu | -196.407117 | -196.425986 | 39.715 |
| $Cu_2O$ | -408.822668 | -408.855849 | 69.836 |
| PenicillinV | -210.474454 | -210.553083 | 165.490 |
| PenicillinV⋯Cu | -406.896287 | -406.982259 | 180.944 |
| PenicillinV⋯$Cu_2O$ | -619.389147 | -619.476349 | 183.532 |
| Streptomycin | -402.013794 | -402.124506 | 233.014 |
| Streptomycin⋯Cu | -598.453445 | -598.572221 | 249.986 |
| Streptomycin⋯$Cu_2O$ | -810.861878 | -810.990852 | 271.447 |

**Table S2.** Calculated values of Gibbs free energies of reaction (ΔG) for various hypothetical supramolecular association processes (in kcal/mol).

| Supramolecular association process | ΔG |
|---|---|
| PenicillinV + Cu → PenicillinV⋯Cu | -2.0 |
| PenicillinV + $Cu_2O$ → PenicillinV⋯$Cu_2O$ | -42.3 |
| Streptomycin + Cu → Streptomycin⋯Cu | -13.6 |
| Streptomycin + $Cu_2O$ → Streptomycin⋯$Cu_2O$ | -6.6 |

**Machine learning**

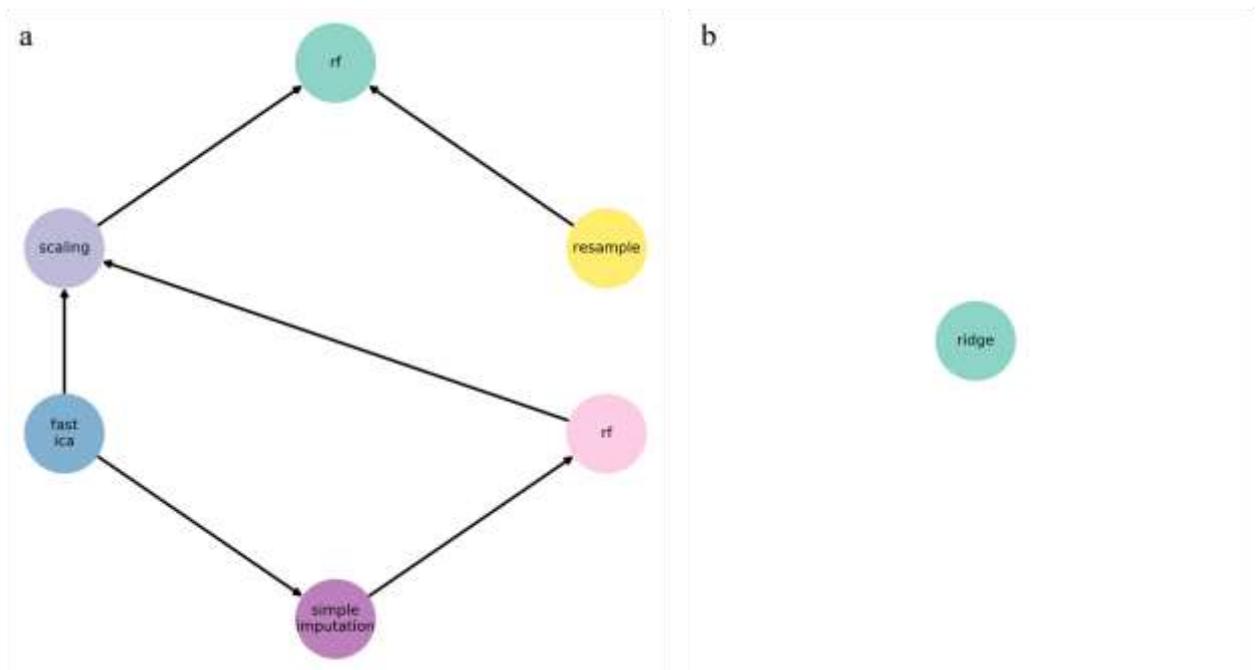

**Figure S2.** Pipeline for 5 antibioitics dataset based on Fedot automl; a- pipeline structure for classification, b - pipeline structure for regression.

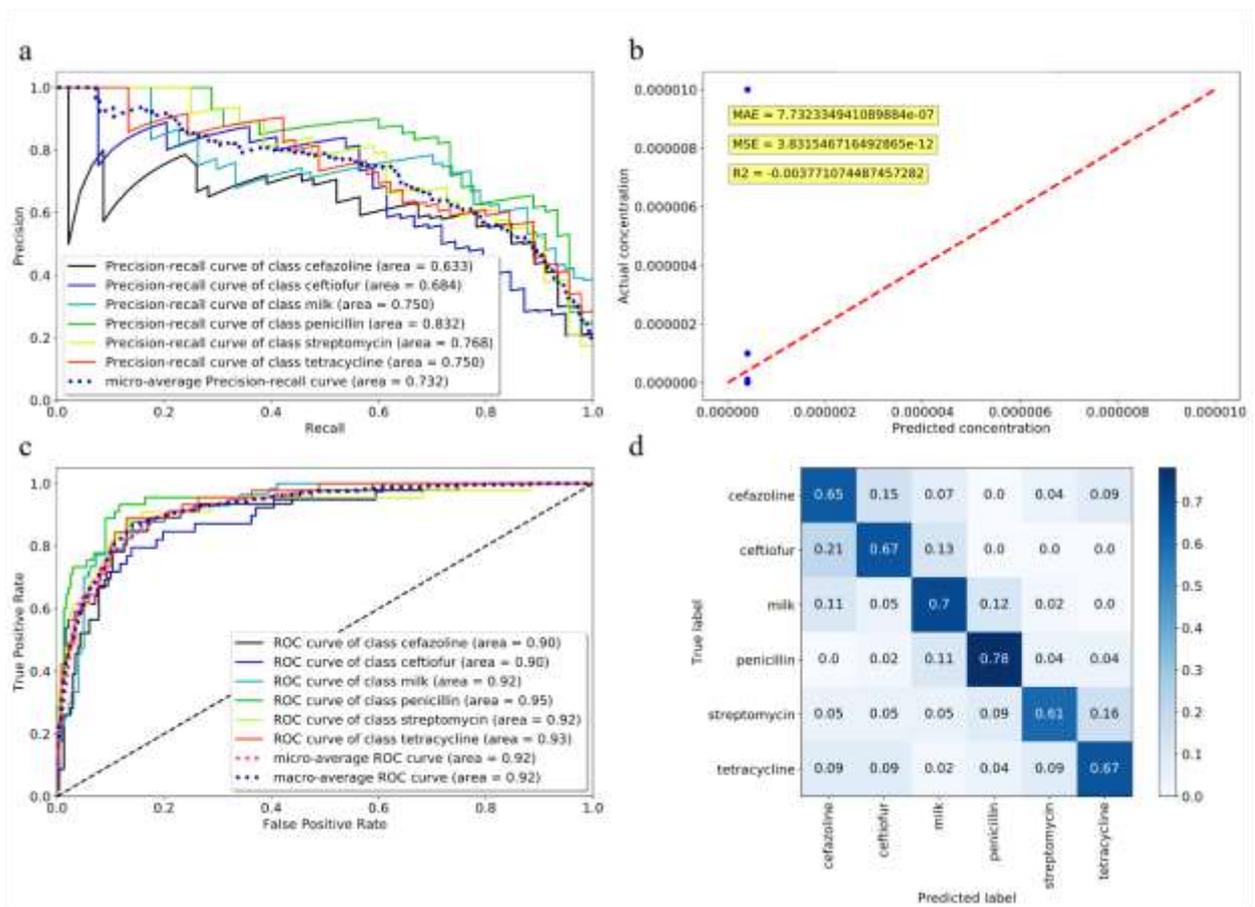

**Figure S3.** Validation results for pipeline with 5 antibiotics based on Fedot automl; a - precision-recall curves for each class; b - regression validation on test data; c - ROC curves for each class; d - normalized confusion matrix on test data.

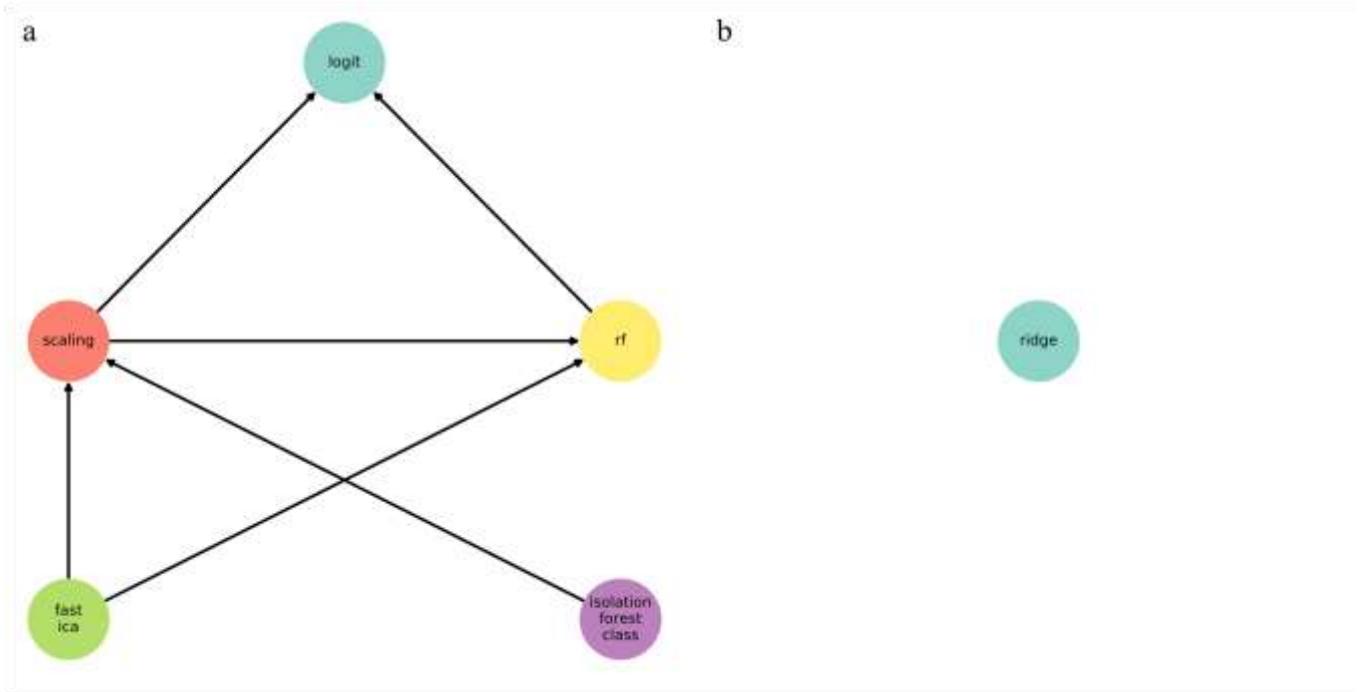

**Figure S4.** Pipeline for 16 antibioitics dataset based on Fedot automl; a - pipeline structure for classification; b - pipeline structure for regression.

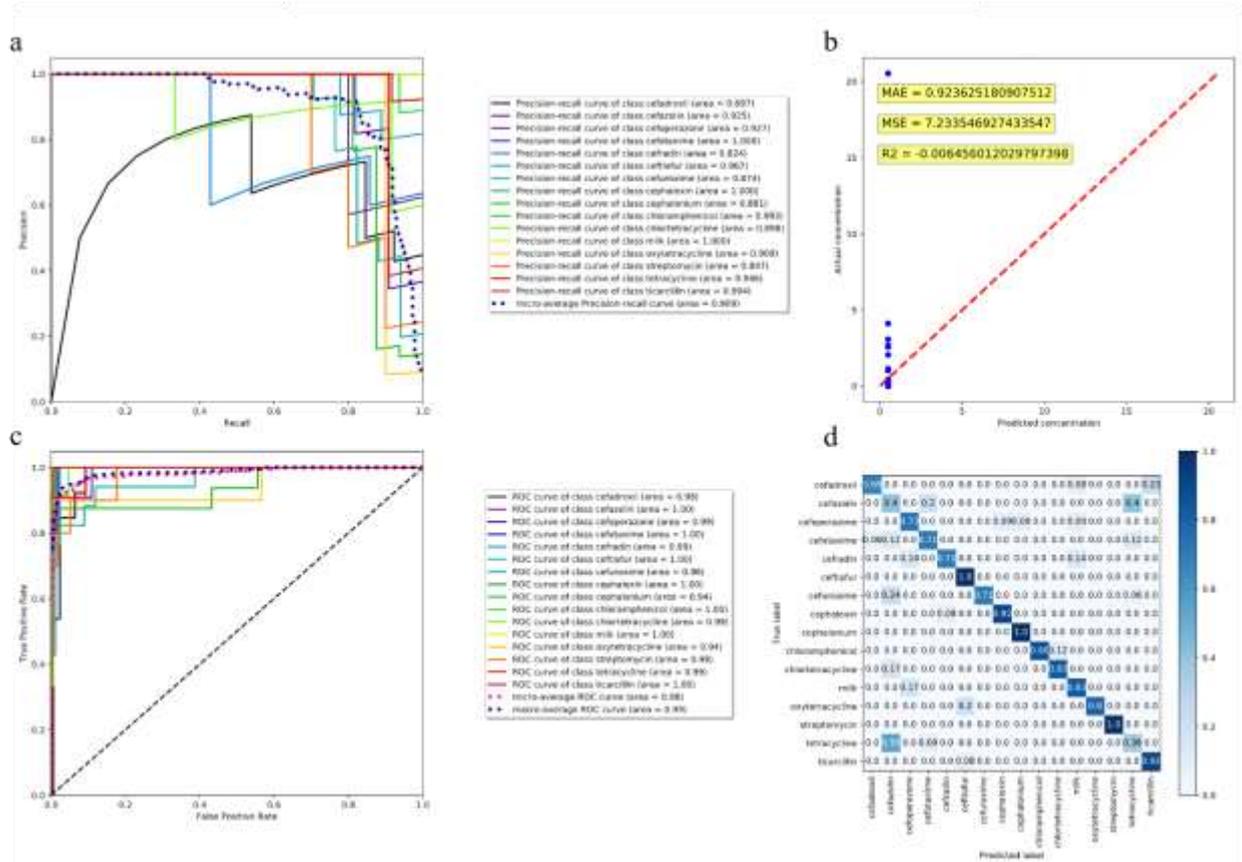

**Figure S5.** Validation results for pipeline with 16 antibiotics based on Fedot automl; a - precision-recall curves for each class; b - regression validation on test data; c - ROC curves for each class; d - normalized confusion matrix on test data.

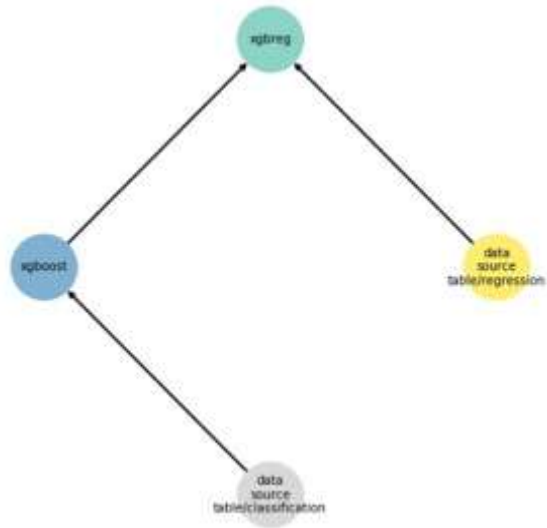

**Figure S6.** Pipeline for classification and regression of 5 antibiotics dataset based on Fedot framework.

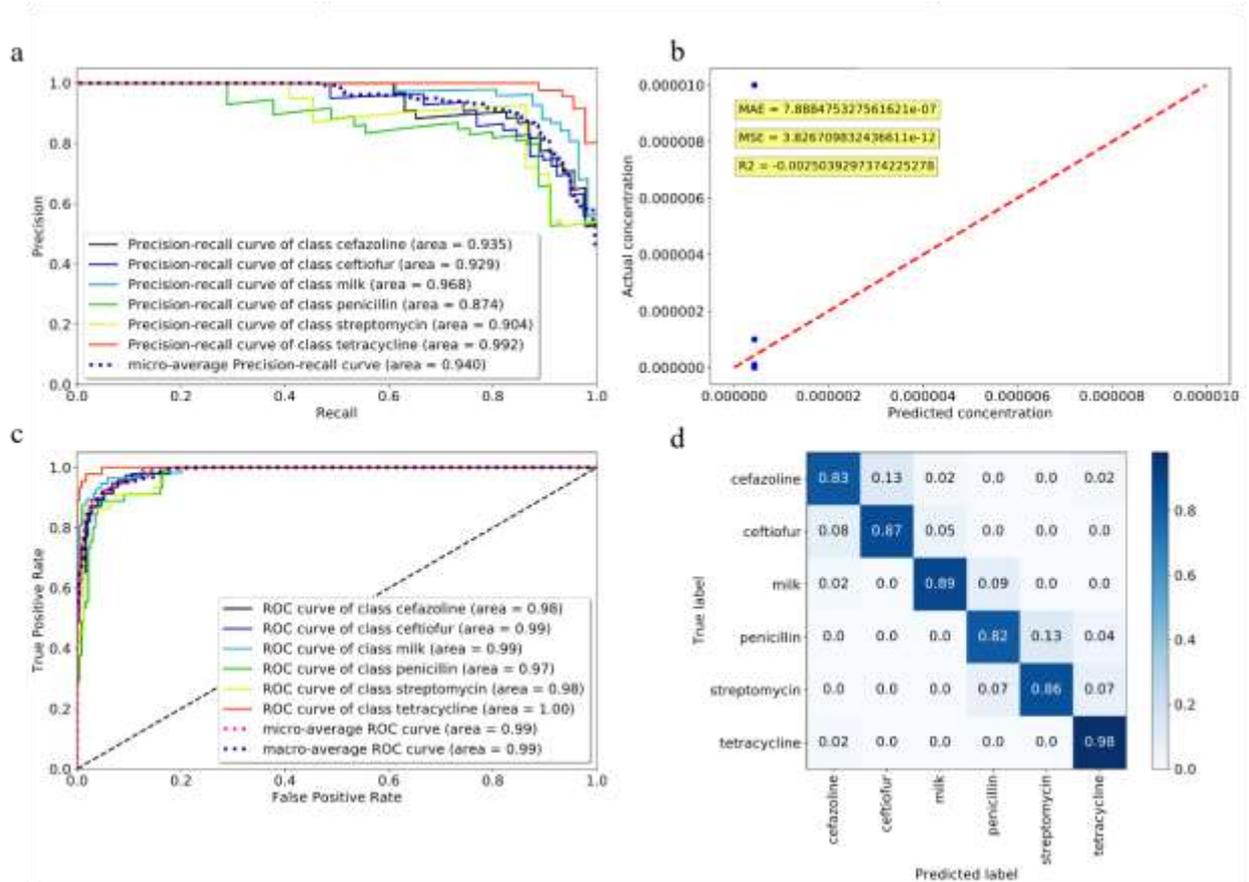

**Figure S7.** Validation results for pipeline with 5 antibiotics based on Fedot framework; a - precision-recall curves for each class; b - regression validation on test data; c - ROC curves for each class; d - normalized confusion matrix on test data.

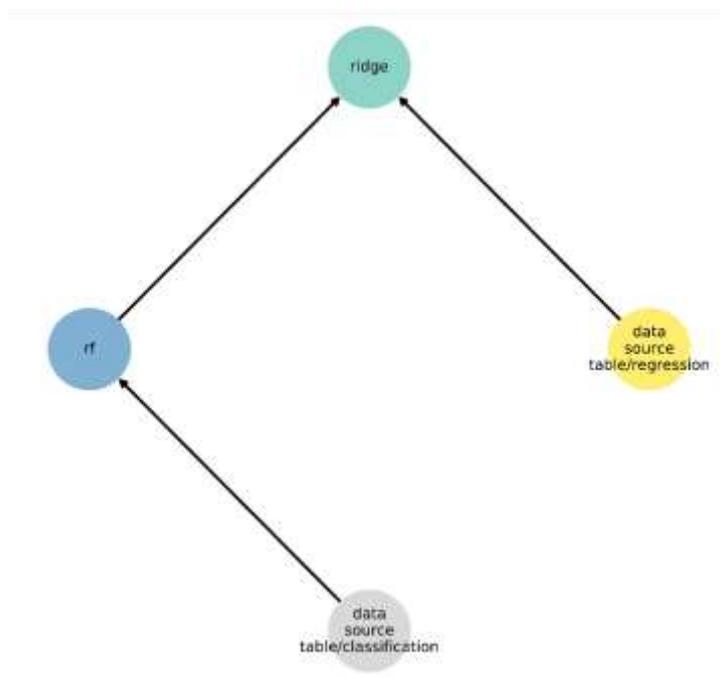

**Figure S8.** Pipeline for classification and regression of 16 antibiotics dataset based on Fedot framework.

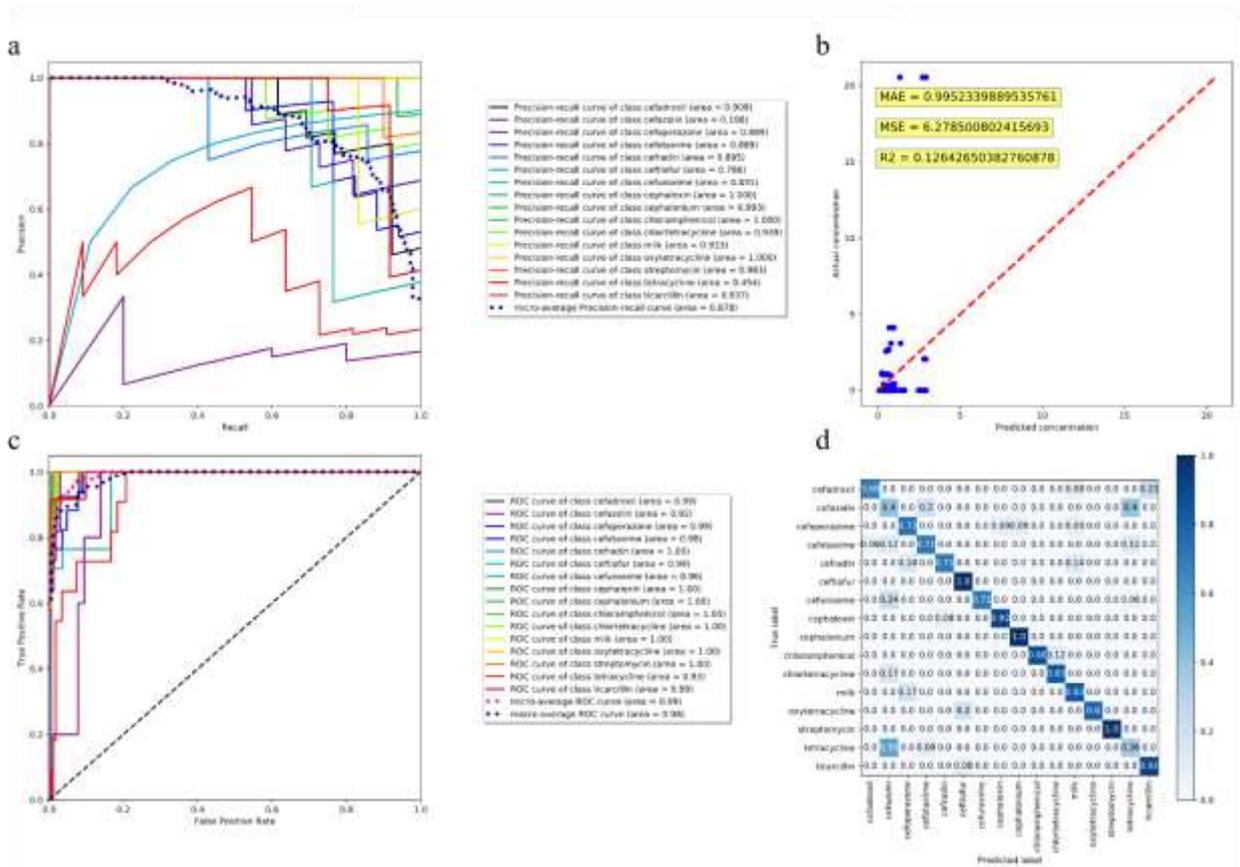

**Figure S9.** Validation results for pipeline with 16 antibiotics based on Fedot framework; a - precision-recall curves for each class; b - regression validation on test data; c - ROC curves for each class; d - normalized confusion matrix on test data.

**Methods**

**Multisensor assembly**

To fabricate the multielectrode sensor, the following materials were taken: single-core copper wires with a cross section of 1.73 and 2.80 mm, nickel wires with a cross section of 0.60 mm, and several carbon cloth strands with a cross section of 1.50 mm each, insulated in heat-shrink tubing (Figure S10).

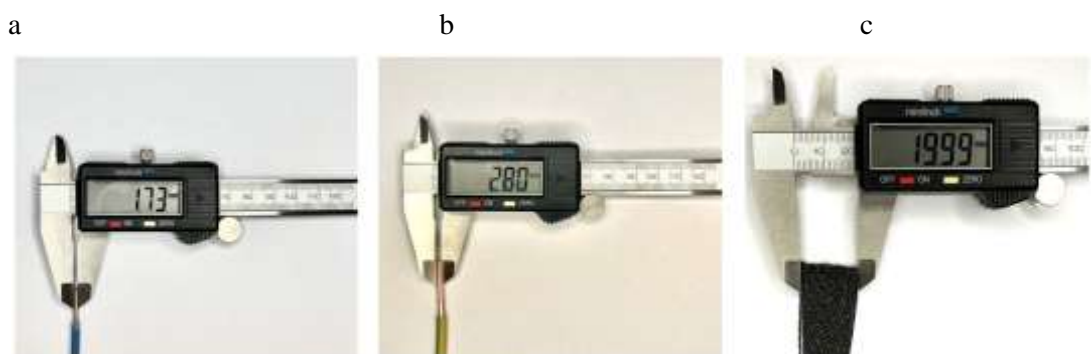

**Figure S10.** Photos of the utilized materials: a – 1.73 mm copper wire in insulating tubing, b – 2.80 mm copper wire, c - a group of carbon fibers.

The final length of the insulated wires (electrodes) was approximately 100 mm.

At the next step, the electrodes, i. e. 3 copper wires, 2 nickel wires and 2 carbon fiber strands were sealed in epoxy resin. For this purpose, 10 ml of epoxy resin AquaGlass Citrus were mixed thoroughly with 5 ml of curing agent for 7 minutes. Then the electrodes were bunched together, placed in a silicone mold filled with the solution of epoxy resin, and left for 24 hours (Figure S11).

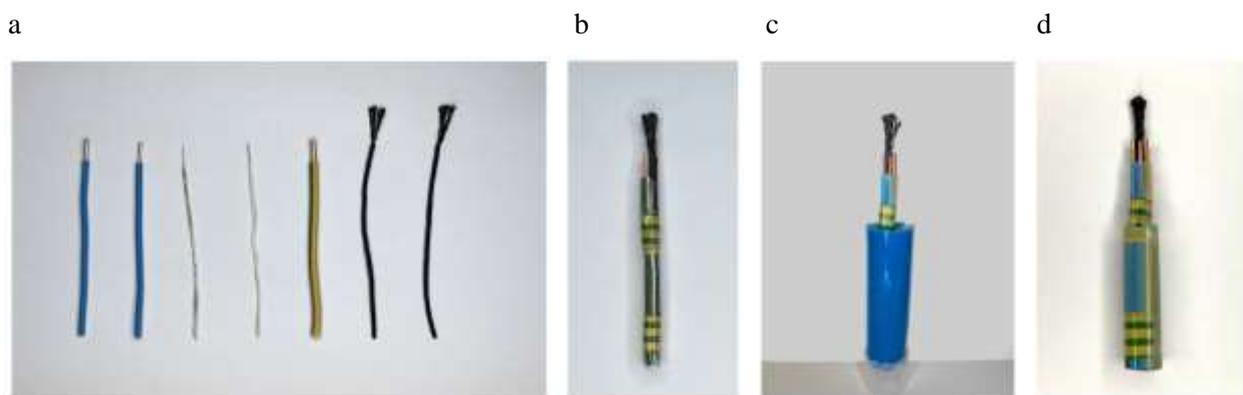

**Figure S11.** Multielectrode sensor assembly: a – pieces of metal wires and carbon fibers, b – electrodes bunched with adhesive tape, c - electrodes in a silicon mold, d – electrodes in solidified epoxy resin.

The solidified body of the sensor was taken out of the mold and carefully polished from its bottom side using Dremel and sandpaper with different grain sizes (from P200 to P2000) to achieve a shiny surface.

Finally, the body of the sensor was placed in a plastic case printed on a 3d printer (Figure S12). Outputs of working electrodes were integrated with a solder. The working and auxiliary electrodes were connected to micro-USB B for convenience.

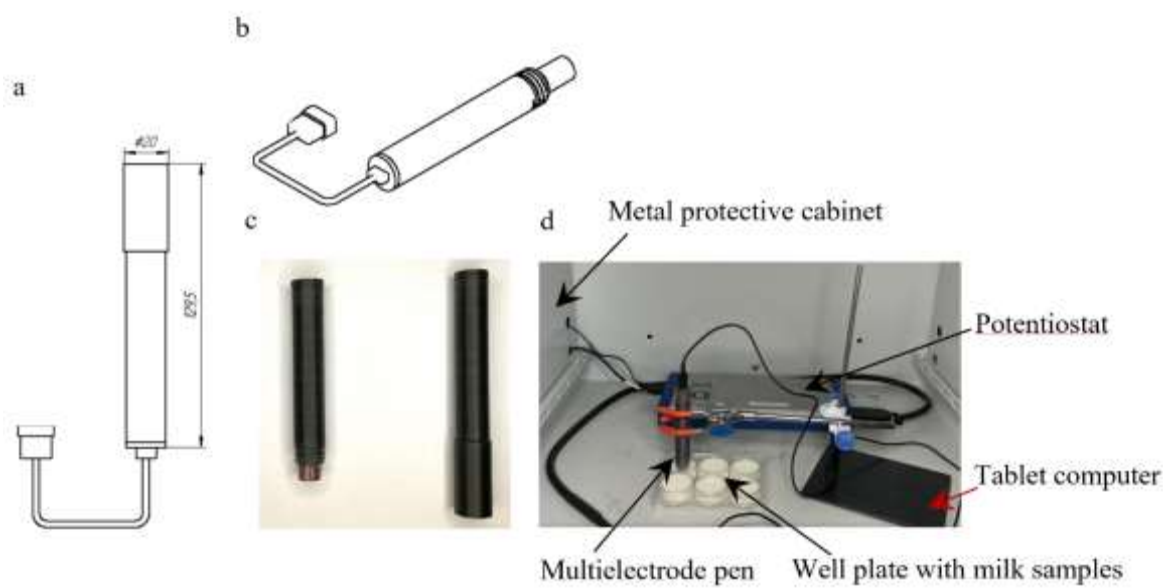

**Figure S12.** Multisensor design and final experimental setup: a - 2d drawing of a sensing pen; b - 3d drawing of a sensing pen; c – photo of a sensing pen in a plastic case; d- experimental electrochemical cell equipped with a potentiostat and electronic device.